# Discrete Element Method Model of Elastic Fiber Uniaxial Compression

Yu Guo[1*], Yanjie Li[2], Qingzhao Liu[2], Zhenhua Li[1], Hanhui Jin[1], Carl Wassgren[3], Jennifer S. Curtis[4]

[1]Department of Engineering Mechanics, Zhejiang University, Hangzhou, 310027, China
[2]School of Technology, Beijing Forestry University, Beijing 100083, China
[3]School of Mechanical Engineering, Purdue University, West Lafayette, IN 47907, USA
[4]Department of Chemical Engineering, University of California Davis, Davis, CA 95616, USA
[*]Corresponding author: yguo@zju.edu.cn

**Abstract**

A flexible fiber model based on the discrete element method (DEM) is presented and validated for the simulation of uniaxial compression of flexible fibers in a cylindrical container. It is found that the contact force models in the DEM simulations have a significant impact on compressive forces exerted on the fiber bed. Only when the geometry-dependent normal contact force model and the static friction model are employed, the simulation results are in good agreement with experimental results. Systematic simulation studies show that the compressive force initially increases and eventually saturates with an increase in the fiber-fiber friction coefficient, and the fiber-fiber contact forces follow a similar trend. The compressive force and lateral shear-to-normal stress ratio increase linearly with increasing fiber-wall friction coefficient. In uniaxial compression of frictional fibers, more static friction contacts occur than dynamic friction contacts with static friction becoming more predominant as the fiber-fiber friction coefficient increases.

## 1. Introduction

Processing of fibrous granular materials is required in various industries, such as those found in renewable energy production with biomass, agricultural crop harvesting, textile production, papermaking, and fiber-reinforced composite production. The poor flowability and complex mechanical properties of flexible fibers results in difficulties during processing. Thus, the development of a good model for fiber dynamics is desirable for improving process and product performance. In recent years, computational simulation has become an important tool for investigating the mechanical behavior of granular materials, including flexible fibers. Compared to designs based on experimental approaches, designs utilizing computational simulations are usually less costly, less time-consuming, and more informative (e.g., micro-scale and transient information can be obtained).

Computational models of flexible fibers typically consist of a chain of elements, which can be spheres, prolate spheroids, cylindrical rods, or sphero-cylinders (Guo and Curtis, 2015). The adjacent elements are connected by virtual bonds or ball-socket joints. The bending, axial extension/compression, and twisting of bonds results in the deformation of a fiber. The bonded-sphere fiber model is described in detail in Guo et al. (2013) and it is found that the critical time step is determined by the length of a virtual bond. Effeindzourou et al. (2016) and Kunhappan et al. (2017) proposed a general scheme for the modeling of deformable structures, which is applicable in the simulation of flexible fibers. In their work, a fiber is treated as a chain of sphero-cylinders (Kunhappan et al., 2017). Considerable simulation work has been done on fiber suspensions in a liquid (Lindstrom and Uesaka, 2007; Wu and Aidun, 2010; du Roure et al., 2019). The orientation, distribution, and deformation of large aspect ratio fibers in suspensions were systematically studied. The concentration of fibers in the suspensions is usually small with fiber-fiber collisions occurring infrequently. Dense systems of fibers have also been numerically investigated. Langston et al. (2015) simulated the packing of flexible fibers in a cylindrical vessel, examining the sensitivity of solid volume fraction to simulation parameters such as initial particle velocity, particle elastic stiffness, and particle friction coefficient. Grof et al. (2007) and Guo et al. (2017) employed the bonded-sphere fiber model in studies of the breakage of elongated particles in a densely-packed bed subject to compression and agitation. In these works, a breakage criterion is introduced to determine when virtual bonds are disconnected. Dense shear flows of flexible fibers were investigated using the bonded-sphere fiber model by Guo et al. (2015; 2019) where the effects of inter-fiber friction, fiber surface roughness, fiber flexibility, and fiber aspect ratio on the shear stress were discussed. In most simulation work, fibers are assumed to be elastic and the plasticity of fibers has been ignored. In the series work by Leblicq et al. (2015; 2016a,b), a flexible fiber model based on connected sphero-cylinders was developed to simulate a collection of bendable crop stems subject to uniaxial compression. In their simulations, crop stems can undergo plastic deformation and a novel database approach was utilized to determine stem bending and stem-stem contact in which the bending stiffness and contact stiffness are obtained from look-up tables of crop stem measurements. Thus, this fiber model is unique to the modeling of specific crop stems. Guo et al. (2018) developed an elasto-plastic fiber model to simulate the permanent plastic deformation of materials such as metal wires.

In the present study, a flexible fiber model based on bonded sphero-cylinders is first described in detail. Three contact types exist between two sphero-cylinders: hemisphere-hemisphere contact, hemisphere-cylinder contact, and cylinder-cylinder contact. In previous flexible fiber simulations, the same contact force model was used for different contact types (Lindstrom and Uesaka, 2007; Effeindzourou et al., 2016; Kunhappan et al., 2017). However, different contact types lead to different sizes and geometries of the contact areas, inevitably causing differences in contact forces. Thus, the same contact force model for all contact types is not accurate in principle. In the present simulations, a specific contact force model is assigned to each contact type, according to the work of Kidokoro et al. (2015). Two-fiber contact experiments are conducted to calibrate a scaling parameter in the contact force models. Next, uniaxial compression tests of a collection of fibers are performed to validate the numerical models. Using the verified fiber model, the effects of fiber-fiber friction and fiber-wall friction on the mechanical response of fiber packings subject to compression are investigated.

## 2. Flexible fiber model

A flexible fiber model is developed based on the DEM method. In this model, a fiber is formed by connecting several identical sphero-cylinders, with a sphero-cylinder consisting of a cylinder capped by a hemisphere at each end as illustrated in Figure 1a (the figure shows 2D sketches but the real models are 3D). Sphero-cylinders are connected at the centers of the hemispheres and the connected points are defined as nodes. The translational and rotational motion of a node sphere is governed by Newton's second law of motion:

$$m_i \frac{d\mathbf{v}_i}{dt} = \mathbf{F}_i^c + \mathbf{F}_i^b + m_i g \quad (1)$$

and

$$J_i \frac{d\boldsymbol{\omega}_i}{dt} = \mathbf{M}_i^c + \mathbf{M}_i^b \quad (2)$$

in which $\mathbf{v}_i$ and $\boldsymbol{\omega}_i$ are the translational and angular velocity vectors, respectively, of node sphere *i* with mass $m_i$ and moment of inertia $J_i$. The translational movement of the node sphere is driven by the contact force $\mathbf{F}_i^c$, the bond force $\mathbf{F}_i^b$, and the gravitational force $m_i \mathbf{g}$. Rotational movement is induced by the moments $\mathbf{M}_i^c$ and $\mathbf{M}_i^b$ due to the contact forces and the bond forces, respectively. The bond forces and bond moments are functions of bond deformation, which is described by the relative displacements between two connected node spheres. Hence, the normal and tangential bond forces $F_n^b$ and $F_t^b$ are expressed as linear functions of normal and tangential displacements $\Delta_n^b$ and $\Delta_t^b$, respectively,

$$F_n^b = \frac{E_b A}{l_b} \Delta_n^b = K_n^b \Delta_n^b, \quad (3)$$

and

$$F_t^b = \frac{G_b A}{l_b} \Delta_t^b = K_t^b \Delta_t^b. \quad (4)$$

The bond twisting moment $M_{\text{twist}}^b$ and bond bending moment $M_{\text{bend}}^b$ are computed incrementally based on the relative twisting angular velocity $\dot{\theta}_{\text{twist}}$ and relative bending angular velocity $\dot{\theta}_{\text{bend}}$ between two connected node spheres,

$$dM_{\text{twist}}^{\text{b}} = \frac{G_{\text{b}}I_{\text{p}}}{l_{\text{b}}}\dot{\theta}_{\text{twist}}dt = K_{\text{twist}}^{\text{b}}\dot{\theta}_{\text{twist}}dt \tag{5}$$

and

$$dM_{\text{bend}}^{\text{b}} = \frac{E_{\text{b}}I}{l_{\text{b}}}\dot{\theta}_{\text{bend}}dt = K_{\text{bend}}^{\text{b}}\dot{\theta}_{\text{bend}}dt \tag{6}$$

In Eqs.(3) - (6), $E_{\text{b}}$ and $G_{\text{b}}$ ($G_{\text{b}} = \frac{E_{\text{b}}}{2(1+\nu_{\text{b}})}$ where $\nu_{\text{b}}$ is the Poisson's ratio of the bond) are the elastic modulus and shear modulus, respectively, of the bond material; $A$ and $l_{\text{b}}$ are the cross-sectional area and length, respectively, of a bond; $I = \pi r^4/4$ is the area moment of inertia; $I_{\text{p}} = \pi r^4/2$ is the polar area moment of inertia; $r$ is the radius of the fiber; and $dt$ is the time step. An illustration of bond forces and moments acting on a node sphere is shown in Figure 1b.

The kinetic energy can be dissipated through deformation and vibration of the flexible fibers. This type of kinetic energy loss is implemented through bond damping forces and moments:

$$F_{\text{n}}^{\text{bd}} = \beta_{\text{b}}\sqrt{2m_i K_{\text{n}}^{\text{b}}}\, v_{\text{n}}^{\text{r}}, \tag{7}$$

$$F_{\text{t}}^{\text{bd}} = \beta_{\text{b}}\sqrt{2m_i K_{\text{t}}^{\text{b}}}\, v_{\text{t}}^{\text{r}}, \tag{8}$$

$$M_{\text{twist}}^{\text{bd}} = \beta_{\text{b}}\sqrt{2J_i K_{\text{twist}}^{\text{b}}}\, \dot{\theta}_{\text{twist}}, \tag{9}$$

and

$$M_{\text{bend}}^{\text{bd}} = \beta_{\text{b}}\sqrt{2J_i K_{\text{bend}}^{\text{b}}}\, \dot{\theta}_{\text{bend}}, \tag{10}$$

where $K_{\text{n}}^{\text{b}}$, $K_{\text{t}}^{\text{b}}$, $K_{\text{twist}}^{\text{b}}$, and $K_{\text{bend}}^{\text{b}}$ represent the normal, shear, twisting, and bending stiffnesses, respectively, of the bond, as defined in Eqs. (3) - (6). The symbols, $v_{\text{n}}^{\text{r}}$, $v_{\text{t}}^{\text{r}}$, $\dot{\theta}_{\text{twist}}$, and $\dot{\theta}_{\text{bend}}$, represent the relative normal velocity, tangential velocity, twisting angular velocity, and bending angular velocity, respectively, between two bonded node spheres of mass, $m_i$ and moment of inertia, $J_i$. The kinetic energy dissipation rate due to the deformation and vibration of the flexible fibers is determined by the bond damping coefficient, $\beta_b$. The larger $\beta_b$, the faster the energy is dissipated.

The normal component $F_n^c$ and tangential component $F_t^c$ of the contact force $\mathbf{F}^c$ exerted on a sphero-cylinder element are linearly distributed to the two node spheres of the element, as shown in Figure 1c. The normal and tangential components of the contact force on node sphere 1 can be expressed as

$$F_n^1 = \frac{\lambda_2}{\lambda_1+\lambda_2} F_n^c , \qquad (11)$$

and

$$F_t^1 = \frac{\lambda_2}{\lambda_1+\lambda_2} F_t^c , \qquad (12)$$

in which $\lambda_1$ and $\lambda_2$ are the distances between the contact point and the tangent points on the node spheres 1 and 2, respectively. Similarly, the force components on node sphere 2 have the expressions

$$F_n^2 = \frac{\lambda_1}{\lambda_1+\lambda_2} F_n^c , \qquad (13)$$

and

$$F_t^2 = \frac{\lambda_1}{\lambda_1+\lambda_2} F_t^c . \qquad (14)$$

The mechanical behavior of a single fiber is validated by simulating cantilever beam bending of a fiber as shown in Figure 2a. A fiber of aspect ratio $AR$ = 25 is fixed on one end with no translation and rotation of the node sphere (highlighted by the red circles in Figure 2a). A load $F_t^0$ is applied to the center of the node sphere at the free end causing the bending of the fiber and the deflection of the free end, $y_0$, as shown in Figure 2a. The normalized cantilever bending deflection of the free end, $y_0/L_c$, is plotted as a function of the normalized load, $(F_t^0 L_c^2)/(E_b I)$, in Figure 2b. In the present numerical model, a fiber can be discretized using various numbers of nodes, $N_n$. A larger number of nodes allows a numerical fiber model to have a better prediction of a continuum fiber. Figure 2b shows that the data with various numbers of nodes $N_n$ = 7, 11, and 21 collapse on the same master curve, which agrees very well with the theoretical prediction of large deformation of a thin beam (Gere and Timoshenko, 1987; Belendez et al., 2002). Thus, it is evident that using more than seven nodes can give a valid simulation of fiber bending behavior.

## 3. Contact force models

To determine if two sphero-cylinders are in contact, the shortest distance $L_{sh}$ between the major axes of two elements is first calculated. The overlap between two sphero-cylinder elements of radius $r$ can be calculated as

$$\delta_n = 2 \cdot r - L_{sh} \ . \tag{15}$$

If $\delta_n$ is greater than zero, the two sphero-cylinders are in contact. The normal contact force model of sphero-cylinders proposed by Kidokoro et al. (2015) is modified and employed in the present work. For a contact between two sphero-cylinders, three different contact types exist: hemisphere-hemisphere contact, hemisphere-cylinder contact, and cylinder-cylinder contact, as shown in Figure 3. Contact force models are different for the different contact types due to the difference in the size and shape of the contact area. The normal contact force for a hemisphere-hemisphere contact (Figure 3a) is described by the Hertzian model as

$$F_n = \frac{\sqrt{2r}}{3} \frac{E_c}{(1-v^2)} \delta_n^{\frac{3}{2}} \ , \tag{16}$$

in which $E_c$ is the elastic modulus of the fiber material at the contact points and $v$ is the Poisson's ratio of the material. The normal contact force for a hemisphere-cylinder contact (Figure 3b) has the form

$$F_n = \sqrt{\frac{8r}{27}} \alpha^{-\frac{3}{2}} \frac{E_c}{(1-v^2)} \delta_n^{\frac{3}{2}} \ , \tag{17}$$

where the parameter $\alpha$ depends on the shape of contact area and is determined to be 0.974 by Kidokoro et al. (2015).

For a parallel cylinder-cylinder contact (Figure 3c), the normal contact force is linearly proportional to the overlap $\delta_n$,

$$F_n = \frac{\kappa \pi L_c}{2\left(1.8864 + \ln\frac{L_c}{2b}\right)} \frac{E_c}{(1-v^2)} \delta_n \ , \tag{18}$$

in which $L_c$ is the length of the contact area along the major axis, $b$ is the width of contact area, $b = \sqrt{2r\delta_n}$, and $\kappa$ is a constant, which is experimentally determined.

For a skewed cylinder-cylinder contact (Figure 3d), the normal contact force depends on the angle between the two major axes, $\theta$. A bilinear model shown in Figure 4 is a simplified version of the model proposed by Kidokoro et al. (2015). In this model, $F_n^{min}$, which is the normal contact force when two sphero-cylinders are perpendicular to each other (i.e., $\theta = \pi/2$), has the expression

$$F_n^{\min} = \frac{2}{3}\kappa\sqrt{r}\alpha^{-\frac{3}{2}}\frac{E_c}{(1-\nu^2)}\delta_n^{\frac{3}{2}} . \tag{19}$$

When $\theta$ is equal to zero, the contact becomes the parallel cylinder-cylinder contact. Thus, $F_n^{\max}$ in Figure 4 has the same expression as Eq. (18),

$$F_n^{\max} = \frac{\kappa\pi L_c}{2\left(1.8864+\ln\frac{L_c}{2b}\right)}\frac{E_c}{(1-\nu^2)}\delta_n , \tag{20}$$

in which $L_c$ is the length of the overlap after rotating one of the sphero-cylinders to the parallel position with the other one from the acute angle direction, and the rotation is conducted about the current contact point (as illustrated in Figure 3d). Therefore, the normal contact force model for a skewed cylinder-cylinder contact can be described as

$$F_n = \begin{cases} F_n^{\max} - \frac{10(F_n^{\max}-2F_n^{\min})}{\pi}\theta, & 0 \leq \theta < 0.1\pi \\ 2F_n^{\min} - \frac{5F_n^{\min}}{2\pi}(\theta - 0.1\pi), & 0.1\pi \leq \theta < 0.5\pi \end{cases} \tag{21}$$

For the contact between a sphero-cylinder element and a flat wall, two types of contacts exist: parallel contact and oblique contact. For the parallel contact in which the major axis of the sphero-cylinder element is parallel to the wall, the contact force is written as

$$F_n = \frac{\kappa\pi L_c}{\left(1.8864+\ln\frac{L_c}{2b}\right)}\frac{E_c E_w}{(1-\nu_w^2)E_c+(1-\nu^2)E_w}\delta_n , \tag{22}$$

in which the width of contact area $b$ has the expression $b = 2\sqrt{r\delta_n}$ and the length of contact area $L_c$ is equal to the length of the cylinder part of the element. The symbols $E_w$ and $\nu_w$ represent the elastic modulus and Poisson's ratio of the wall material. Contact between the end hemisphere and the wall has a normal contact force given by

$$F_n = \frac{4\sqrt{r}}{3}\frac{E_c E_w}{(1-\nu_w^2)E_c+(1-\nu^2)E_w}\delta_n^{\frac{3}{2}} . \tag{23}$$

Note that although the overlap remains constant when the contact type changes, e.g., from hemisphere-hemisphere to hemisphere-cylinder, the normal force changes abruptly due to the change in the contact geometry. Kumar et al. (2018) developed transition functions to smoothly transition between contact force types; however, these were not employed in the present work. For the simulations performed here, force discontinuities due to contact type transitions were not large enough to significantly perturb the fiber bed and, thus, were not needed to ensure bed stability.

Two tangential force models have been employed: the Coulombic sliding friction model and the Mindlin model. The Coulombic sliding friction model accounts for 'sliding' only and the magnitude of the tangential force is equal to the product of the dynamic friction coefficient $\mu$ and normal contact force $F_\text{n}$:

$$\boldsymbol{F}_\text{t} = -\mu F_\text{n} \frac{\boldsymbol{v}_c^t}{|\boldsymbol{v}_c^t|} , \qquad (24)$$

in which $\boldsymbol{v}_c^t$ is the tangential component of the relative velocity vector at the contact point. The Mindlin model takes into account both static and sliding processes,

$$\boldsymbol{F}_\text{t} = \begin{cases} \boldsymbol{F}_\text{t}^1 = \boldsymbol{F}_\text{t}^0 + 8G^* a \cdot \boldsymbol{v}_c^t \cdot dt, & |\boldsymbol{F}_\text{t}^1| < \mu F_\text{n} \\ -\mu F_\text{n} \frac{\boldsymbol{F}_\text{t}^1}{|\boldsymbol{F}_\text{t}^1|}, & |\boldsymbol{F}_\text{t}^1| \geq \mu F_\text{n} \end{cases}, \qquad (25)$$

where $\boldsymbol{F}_\text{t}^0$ and $\boldsymbol{F}_\text{t}^1$ are the tangential force vectors in the previous time step and in the current time step, respectively. The static friction coefficient is assumed to be the same to the dynamic friction coefficient, thus both of them are represented by the friction coefficient $\mu$. The parameter $G^*$ is as follows

$$\frac{1}{G^*} = \frac{2-\nu_1}{G_1} + \frac{2-\nu_2}{G_2} , \qquad (26)$$

in which $G_1$ and $G_2$ are the shear moduli of the two sphero-cylinders or a sphero-cylinder and a wall in contact, $\nu_1$ and $\nu_2$ are the corresponding Poisson's ratios, and the shear modulus and elastic modulus have the correlation $G_i = E_i/(2(1+\nu_i))$. The effective radius of contact $a$ is defined as $a = \sqrt{2r\delta_\text{n}}/2$, and $\boldsymbol{v}_c^t dt$ represents the incremental tangential displacement in the present time step.

In the simulations, contact damping forces are added to dissipate the kinetic energy during collisions. The normal contact damping force is expressed as

$$F_\text{dn}^\text{c} = -c\beta_\text{c}\sqrt{2m_\text{i} K_\text{n}^\text{c}}\, v_\text{c}^\text{n} , \qquad (27)$$

in which $\beta_\text{c}$ is the contact damping coefficient, which determines the collisional dissipation rate. The normal contact stiffness $K_\text{n}^\text{c}$ is defined as $K_\text{n}^\text{c} = dF_\text{n}/d\delta_\text{n}$. The quantity $c$ is equal to one if the normal contact force $F_\text{n}$ is proportional to $\delta_\text{n}$ and $c$ is equal to $\sqrt{5/6}$ if the normal contact force $F_\text{n}$ is proportional to $\delta_\text{n}^{\frac{3}{2}}$. The parameter $v_\text{c}^\text{n}$ is the relative normal velocity at the contact point and the negative sign indicates the direction of the damping force vector is opposite to that of the relative velocity. The tangential contact damping force, which exists only in the static friction process in the Mindlin model, is written as

$$F_{dt}^c = -\beta_c \sqrt{2m_i K_t^c}\, v_c^t \,, \tag{28}$$

in which the tangential contact stiffness $K_t^c$ is expressed as $K_t^c = 8G^* a$.

## 4. Modeling uniaxial compression

The mechanical response of an assembly of fibers subject to compression is important for the handling and processing of fibrous materials in engineering practice. Using the present fiber model, simulations are performed to investigate the uniaxial compression of flexible fibers. Experiments are also conducted for the validation of the numerical fiber model. Cut silicon rubber cords are used in the experiments. To calibrate the contact force model, compression experiments are conducted to measure the contact force between two 100 mm long, 20 mm diameter rubber cords, as shown in Figure 5a. The two cords are perpendicular to each other during the compression. The normal contact force $F_n$ is plotted against the overlap $\delta_n$ in Figure 5b. A scatter bar represents plus and minus a standard deviation from the six replicated tests. The dashed line in the figure is the normal contact force model for the skewed cylinder-cylinder contact, i.e., Eq. (19). It is found that using a value of $\kappa$ = 2.5 in Eq. (19) results in good agreement with the experimental results. Thus, in the present simulations the constant $\kappa$ is set to 2.5 for the normal contact force models (Eqs. (18) - (22)).

The uniaxial compression tests of fiber packings are conducted on a ZwickRoell materials testing machine (model Zwick10KN). As shown in Figure 6a, in the uniaxial compression tests, 550 cut rubber cords (the same material used in the calibration experiment) of diameter $d_f$ = 2.4 mm and length $L_f$ = 60 mm, giving an aspect ratio of 25, are randomly packed in a cylindrical, Perspex container of inner diameter 80 mm. The rubber cords are put into the container by dropping five rubber cords from the top of the container at a time. The Young's modulus and sliding friction coefficient of the rubber cords are measured as $E_b$ = $E_c$ = 6.35 × $10^6$ Pa and $\mu_{ff}$ = 1.4. In the tests, the bed of the rubber cords is compressed by the upper piston. The piston initially moves downwards at a constant speed of 5 mm/s to load the cords until it reaches a distance of about 50 mm from the bottom. Immediately afterwards the piston moves upwards at the same speed to unload the cords. Multiple loading-unloading cycles were conducted. The simulations use a similar set-up and procedure, as shown in Figures 6b and 6c. An upper wall is used to mimic the piston to compress the fibers. The measured properties of the cut rubber cord and the Perspex container are used for the flexible fibers and cylindrical container in the simulations, respectively. The simulation parameters are presented in Table 1.

In Figure 6, the first column shows the experimental results (a, d, and g), the second column shows the DEM simulation results with the Coulombic model of tangential force (b, e, and h), and the third column shows the DEM simulation results with the Mindlin model of tangential force (c, f, and i). The geometry-dependent normal contact force models described in Section 3 are used in the simulations presented in Figure 6. The first row shows the fiber packings before the compression, the second row shows the compressed packings at the lowest position of the piston or the upper wall, and the third row shows the relaxed packings after unloading. The elastic fibers show a rebounding behavior during the unloading process due to the elasticity of the fibers. It is clear that the initial and final bed heights in the DEM simulation with the Coulombic model of tangential force are smaller than those observed from the experiment and the DEM simulation with the Mindlin model of tangential force. This observation indicates that the tangential contact force model has a significant impact on the packing of fibers.

Figure 7 shows the variation of compressive force on the fibers with the solid volume fraction. A scatter bar represents plus and minus a standard deviation from at least three measurements. In the DEM simulations, different combinations of tangential and normal contact force models are investigated. The simplified normal contact force model (simplified $F_n$) is the hemisphere-hemisphere contact model (Eq. (16)), which is used for all three contact types. The geometry-dependent normal contact force model (G-D $F_n$) is the set of models described in Section 3, in which a different normal contact force law is assigned to each contact type. As shown in Figure 7, both normal and tangential force models influence the loading curves. With the Coulombic friction model, a significant compressive force increase occurs at much larger solid volume fractions compared to the Mindlin model. In the Coulombic model (see Eq.(24)), the direction of the tangential contact force is opposite to that of the relative tangential velocity. In the compression process, the direction of the relative tangential velocity at the contact point can change frequently, causing an abrupt change in the direction of the tangential force. In addition, when the relative tangential velocity at the contact point is zero, the tangential force has to be zero according to the Coulombic model. Thus, static friction is not considered in the Coulombic model. The abrupt change of tangential force direction and the lack of static friction force reduce the stability of the fiber bed, leading to smaller bed heights (Figure 6b and 6h) and smaller compressive forces (Figure 7). However, in the Mindlin model, i.e., Eq. 25, the tangential force vector changes gradually by a small incremental vector in each time step, avoiding the abrupt direction change. Static friction is considered so that a tangential force can exist when a zero relative tangential velocity occurs. In conclusion, the Coulombic model of tangential force cannot be employed in simulations of quasi-static fiber packing and compression in which the deformation of fiber bed is slow, and a tangential force model that accounts for static friction, e.g., the Mindlin model, needs to be used for a better prediction of such processes. With a given tangential force model, the simplified normal contact force model leads to a smaller compressive force at large solid volume fractions due to the fact that the simplified model underestimates the magnitudes of contact forces for the cylinder-cylinder type of contacts. It is observed that the simulation results with the Mindlin model and geometry-dependent model for the tangential and normal contact forces, respectively, are in good agreement with the experimental results. Kumar et al. (2018) found that an accurate normal contact force model, which accounts for contact types and contact transitions, is needed to predict more detailed contact information, such as contact area, contact overlap, and contact duration. The present results show that the more accurate normal contact force model that accounts for contact types should be used in the uniaxial compression simulations for a more accurate prediction of compressive forces. Nevertheless, the tangential force model that takes into account static friction is critical in such compression simulations. Hence, the Mindlin model of tangential force and the geometry-dependent model of normal force are employed in the following simulations.

The variation of the compressive force with the solid volume fraction in the first load-unload cycle is shown in Figure 8a for both experiments and simulations. A scatter bar represents plus and minus a standard deviation from at least three tests. The simulation results are consistent with the experimental results. The force curves in the second load-unload cycle are added in Figure 8b in which the scatter bars are removed for visual clarity. Compared to the first cycle, the loading curve in the second cycle becomes lower due to the effect of compaction after the first loading cycle. The unloading curve in the second cycle nearly coincides with the unloading curve in the first cycle, possibly because the compacted fibers at the lowest piston position have very similar structures in the two load-unload cycles. Similar load-unload behaviors are obtained from the experiments and simulations. Figure 9 shows the load-unload cycles with the unloading at various solid volume fractions. It is interesting to observe that the load-unload cycles have similar shapes and the various unload curves tend to converge to the same master curve close to the end of the unloading process. Also, the DEM simulation results are in good agreement with the experimental results.

The effects of fiber-fiber friction coefficient and fiber-wall friction coefficient on the loading curves are shown in Figure 10a and 10b, respectively. All of the simulation cases have nearly the same initial configuration of fibers in the container. At a given solid volume fraction, as fiber-fiber friction coefficient increases, the compressive force $F_\mathrm{p}$ generally increases due to the increased resistance in the relative movement between fibers. With small fiber-fiber friction coefficient, the fibers more easily rearrange themselves, reducing the fiber-fiber contact forces. However, the force eventually saturates as the fiber-fiber friction coefficient is greater than 1.4 (Figure 10a). This result is reasonable considering that if the friction coefficient is sufficiently large then the tangential contact force is independent of the friction coefficient due to static friction. The increase in the fiber-wall dynamic friction also augments the compressive force (Figure 10b) because the wall friction prevents the downward movement of fibers close to the wall. Based on the data obtained in Figure 10a, the empirically-determined polynomial: $(5.7\mu_\mathrm{ff}^3 - 33.8\mu_\mathrm{ff}^2 + 68.5\mu_\mathrm{ff} + 1)$ is obtained to scale $F_\mathrm{p}$. The scaled compressive forces, $F_\mathrm{p}/(5.7\mu_\mathrm{ff}^3 - 33.8\mu_\mathrm{ff}^2 + 68.5\mu_\mathrm{ff} + 1)$, for various fiber-fiber dynamic friction coefficients approximately collapse to a single curve, as shown in Figure 10c. Similarly, based on the data in Figure 10b, the compressive forces $F_\mathrm{p}$ for variations in fiber-wall friction can be scaled by the empirically-determined linear relation: $(0.34\mu_\mathrm{fw} + 1)$, as shown in Figure 10d. Thus, in uniaxial compression, the compressive force exerted on the fiber bed has a simple linear relationship with the fiber-wall dynamic friction coefficient, while it has a much more complex relationship with the fiber-fiber dynamic friction coefficient.

The average normal stress or pressure, $\sigma_{rr}$, exerted on the cylindrical inner wall of the container is plotted against the solid volume fraction in Figure 11a and 11b. The effect of fiber-fiber friction coefficient $\mu_{ff}$ on the normal stress $\sigma_{rr}$ is similar to its effect on the compressive force $F_p$: the normal stress generally increases with increasing $\mu_{ff}$ at a given solid volume fraction, as shown in Figure 11a. The larger fiber-fiber friction coefficients enhance the load-bearing capacity and bulk stiffness of the fiber packing, thus larger compressive force $F_p$ and larger lateral normal stress $\sigma_{rr}$ are required to compress the fiber packing. The effect of fiber-wall friction coefficient $\mu_{fw}$ on the normal stress $\sigma_{rr}$ shows three stages: i) slight change in $\sigma_{rr}$ with $0.1 < \mu_{fw} < 0.6$; ii) an increase in $\sigma_{rr}$ with $0.6 < \mu_{fw} < 1.5$; and iii) saturation of $\sigma_{rr}$ with $\mu_{fw} > 1.5$ (Figure 11b). It is also observed that both the compressive force and normal stress exhibit significant fluctuation with a large fiber-wall friction coefficient $\mu_{fw} = 2.0$. The fluctuation may be because the large fiber-wall frictional force causes the sudden rearrangement of fibers close to the wall, changing the contacts between the fibers and cylindrical wall.

From the simulations, the average shear stress exerted on the inner cylindrical wall of the container, $\sigma_{ry}$, in which r represents the radial direction and y represents the major axis direction of the container, can be obtained. The shear-to-normal stress ratio, $\sigma_{ry}/\sigma_{rr}$, is found to remain nearly constant as the solid volume fraction increases during the compression process. The average stress ratio, $\sigma_{ry}/\sigma_{rr}$, in a compression process is plotted as a function of fiber-fiber friction coefficient and fiber-wall friction coefficient in Figure 12. The stress ratio shows a limited change with the fiber-fiber friction coefficient (Figure 12a), while it increases linearly with the increasing fiber-wall friction coefficient (Figure 12b). Therefore, the shear-to-normal stress ratio on the wall depends on the fiber-wall friction coefficient rather than the fiber-fiber friction coefficient. The linear increase of shear-to-normal stress ratio on the wall is consistent with the linear increase of compressive force $F_p$ with increasing fiber-wall friction coefficient. In addition, it is observed that the stress ratio is smaller than the corresponding fiber-wall friction coefficient, which is due to the occurrence of static friction between the fibers and wall in the compression process.

The probability density functions (PDFs) of tangential and normal contact forces between fibers in the compression process from the solid volume fraction 0.2 to 0.55 are plotted in Figure 13. More large tangential contact forces are induced as the fiber-fiber friction coefficient increases, and the PDFs tend to reach an asymptote at large fiber-fiber friction coefficient (Figure 13a). A similar behavior is observed for the PDFs of normal contact forces (Figure 13b). The asymptotic behavior of PDFs of contact forces with increasing $\mu_{ff}$ (Figures 13a and 13b) is consistent with the saturation of compressive forces (Figure 10a). The fiber-wall friction coefficient has an insignificant effect on the PDFs of fiber-fiber contact forces (Figures 13c and 13d), indicating the wall effect on the fiber-fiber contacts is limited. Figure 14 shows the percentage of static friction contacts in the fiber-fiber contacts for various $\mu_{ff}$. A static friction contact is one in which the tangential contact force $F_t$ is less than the product of the fiber-fiber friction coefficient and the normal contact force $\mu_{ff}F_n$, i.e., $F_t < \mu_{ff}F_n$. It can be seen from Figure 14 that the percentage of static friction contacts increases more rapidly with $\mu_{ff}$ when $\mu_{ff} < 1$. For $\mu_{ff} > 1$, over 90% of contacts are static friction contacts, and as a result, the fiber-fiber contact forces tend to be independent of the fiber-fiber friction coefficient under this condition (Figures 13a and 13b). It should be noted that for a given fiber-fiber friction coefficient $\mu_{ff}$, similar percentages of static friction contacts are obtained in a narrow range (0.92-0.93) for fiber-wall friction coefficients varying from 0.1 to 2. Thus, the proportion of static friction contacts is determined by the fiber-fiber friction coefficient, rather than the fiber-wall friction coefficient.

## 5. Conclusions

A bonded sphero-cylinder fiber model is described in detail. In this numerical scheme, fiber-fiber contact considers the different contact scenarios between the sphero-cylinder elements. The normal contact force models proposed by Kidokoro et al. (2015) are modified and utilized to determine the normal contact force for two sphero-cylinder elements in contact. A specific contact force model is implemented for each contact type (hemisphere-hemisphere contact, hemisphere-cylinder contact, and cylinder-cylinder contact) to consider the difference in various contact types. A compression test of two skewed fibers was conducted to calibrate the normal contact force models.

In uniaxial compression simulations of a collection of fibers, both tangential and normal contact force models have an impact on the results. Simulations with a Coulombic friction model or a simplified normal contact force model underestimate the compressive forces. The simulations with both Mindlin tangential contact force model and geometry-dependent normal contact force model are in good agreement with the experimental results. The Coulombic model accounts for sliding friction only while the Mindlin model of tangential force considers both static friction and sliding friction. Thus, static friction has to be taken into account in simulations of densely-packed particle beds that deform slowly. A geometry-dependent normal contact force model considering various contact types is critical for a more accurate simulation of fiber compression at large solid volume fractions.

Systematic studies were performed to investigate the effects of fiber-fiber friction coefficient, $\mu_{ff}$, and fiber-wall friction coefficient, $\mu_{fw}$, on the mechanical response of fibers subject to compression. It is found that the compressive force exerted on the fiber bed initially increases and eventually saturates with an increase in the fiber-fiber friction coefficient, and the fiber-fiber contact forces follow a similar trend. The compressive force increases linearly with increasing fiber-wall friction coefficient, and similarly, the shear-to-normal stress ratio on the cylindrical wall also increases linearly with the fiber-wall friction coefficient. In the uniaxial compression of frictional fibers, more static friction contacts occur than dynamic friction ones for contacting fibers and static friction becomes more predominant as the fiber-fiber friction coefficient increases.

## Acknowledgements

The National Science Foundation of China (Grant NO. 11872333 and NO. 91852205) and the Zhejiang Provincial Natural Science Foundation of China (Grant NO. LR19A020001) are acknowledged for the financial supports.

Table 1. Input parameters for the DEM simulations

| Parameters | Values |
|---|---|
| Flexible fiber diameter, $d_f$ (mm) | 2.4 |
| Flexible fiber length, $L_f$ (mm) | 60 |
| Flexible fiber material density (kg/m$^3$) | 1157.5 |
| Number of nodes in a flexible fiber, $N_n$ (-) | 11 |
| Fiber-fiber contact modulus (i.e., elastic modulus of spheres), $E_c$ (Pa) | $6.35 \times 10^6$ |
| Bond bending modulus (i.e., elastic modulus of bonds), $E_b$ (Pa) | $6.35 \times 10^6$ |
| Poisson's ratio of fibers (-) | 0.3 |
| Fiber-fiber contact damping coefficient, $\beta_c$ (-) | 0.016 |
| Bond damping coefficient, $\beta_b$ (-) | 0.0335 |
| Fiber-fiber friction coefficient, $\mu_{ff}$ (-) | 1.4 |
| Number of fibers (-) | 550 |
| Diameter of Perspex container (mm) | 80 |
| Fiber-wall friction coefficient, $\mu_{fw}$ (-) | 0.6 |
| Elastic modulus of walls (Pa) | $3.2 \times 10^9$ |
| Poisson's ratio of walls (-) | 0.3 |
| Upper wall speed (mm/s) | 5 |
| Time step (s) | $6 \times 10^{-6}$ |

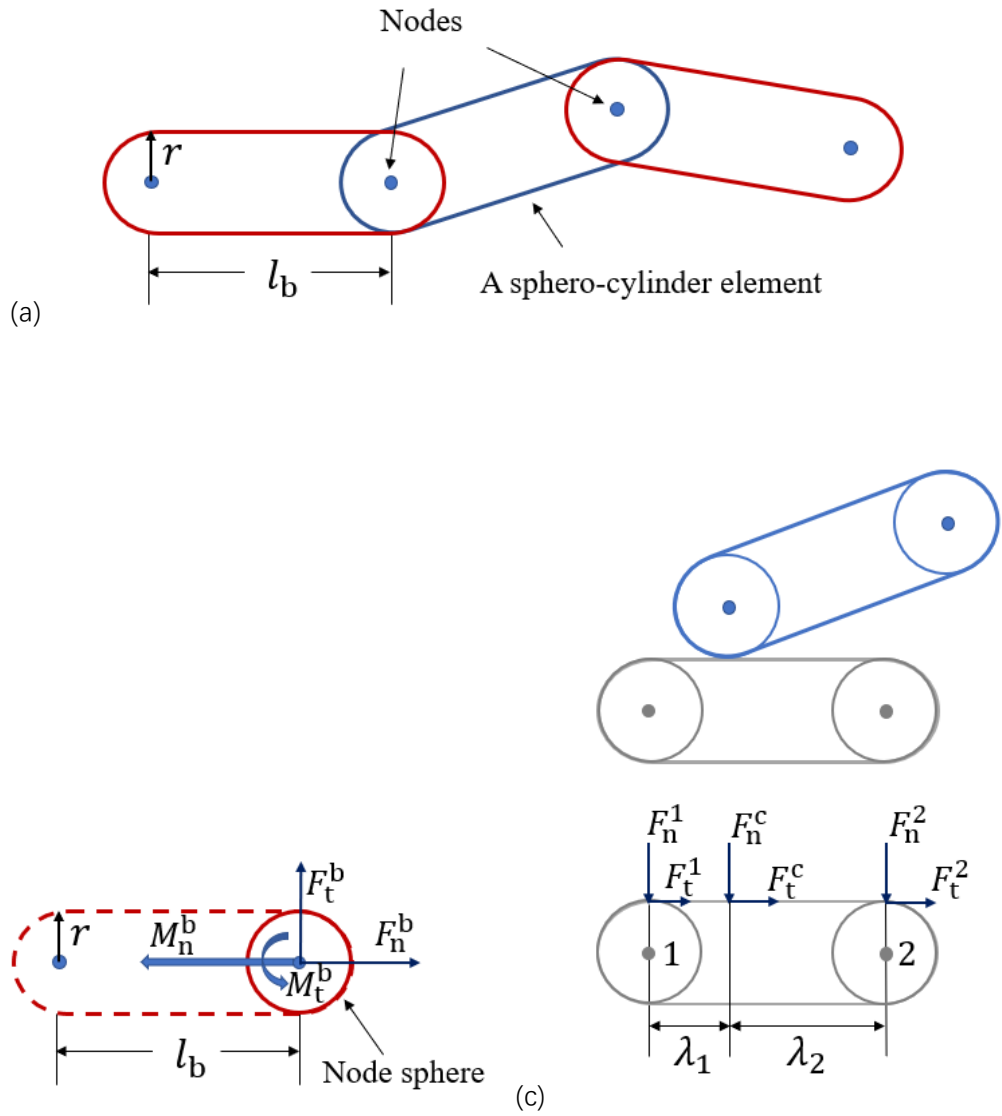

Figure 1. (a) A sketch of the flexible fiber model, (b) bond forces and moments exerted on a node sphere, and (c) contact forces distributed to the two node spheres.

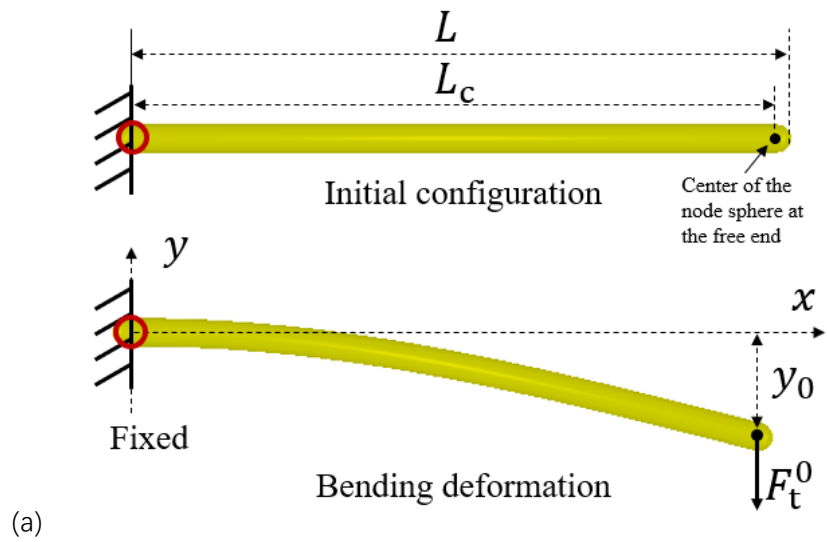

(a)

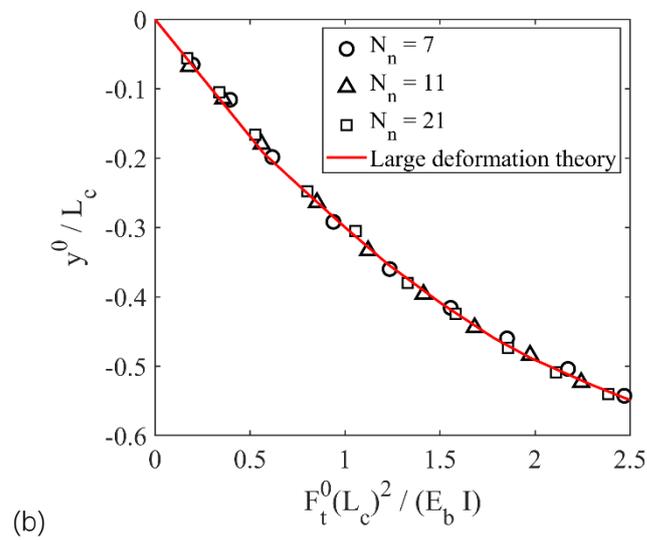

(b)

Figure 2. (a) DEM simulation of cantilever bending of a fiber and (b) normalized cantilever bending deflection of the free end as a function of the normalized load for various numbers of nodes in a fiber. (*AR* = 25)

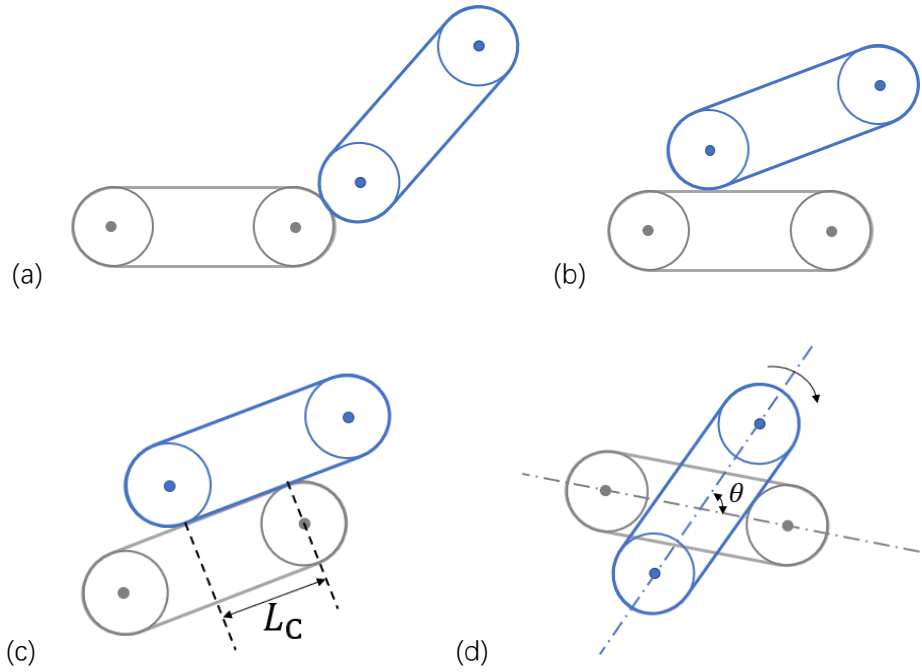

Figure 3. (a) Hemisphere-hemisphere contact, (b) hemisphere-cylinder contact, (c) parallel cylinder-cylinder contact, and (d) skewed cylinder-cylinder contact.

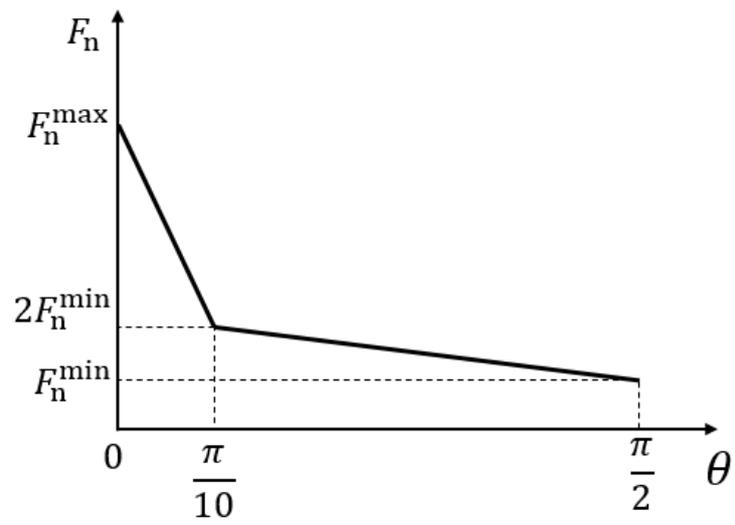

Figure 4. Normal contact force $F_n$ as a function of the angle $\theta$ between the two major axes of the two sphero-cylinder elements in the skewed cylinder-cylinder contact.

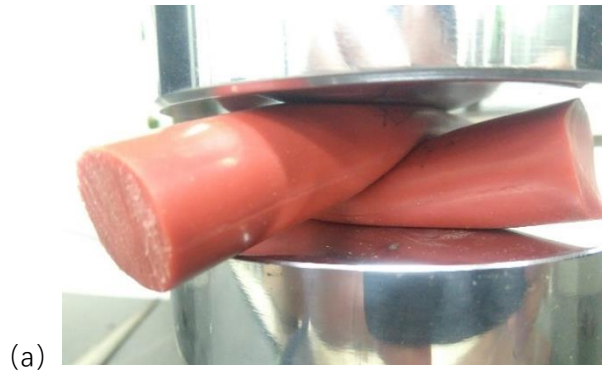

(a)

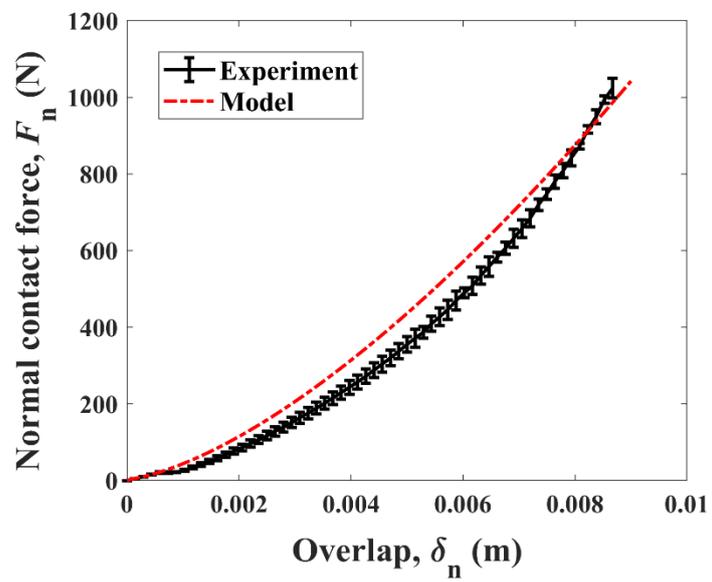

(b)

Figure 5. (a) Experimental measurement of the contact force between two rubber cords that are perpendicular to each other. (b) Normal contact force as a function of overlap between the two contacting cords. (Diameter of the rubber cords is 0.02 m)

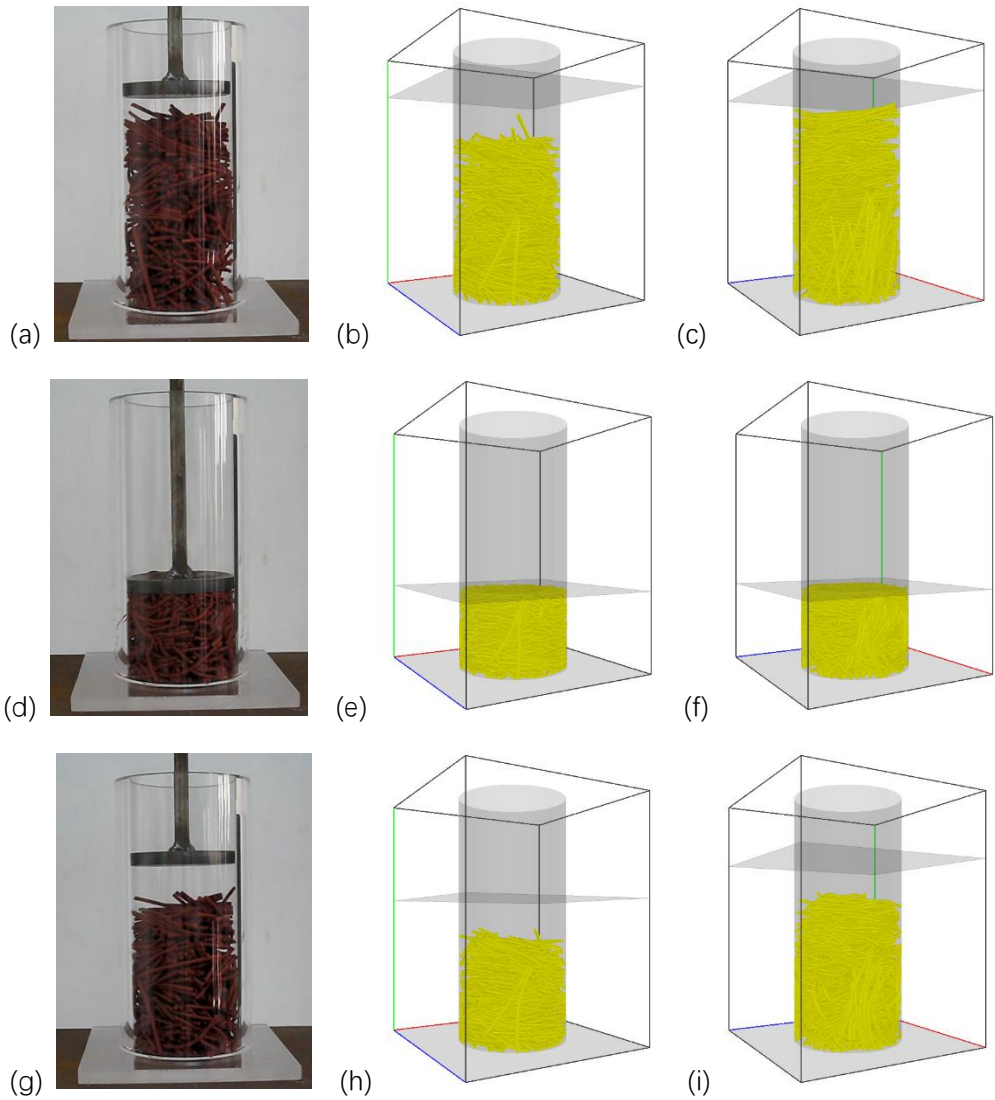

Figure 6. Uniaxial compression of an assembly of rubber cords: The first column shows the experimental results (a, d, and g), the second column shows the DEM simulation results with the Coulombic model of tangential force (b, e, and h), and the third column shows the DEM simulation results with the Mindlin model of tangential force (c, f, and i). The first row shows the fiber packings before the compression, the second row shows the compressed packings at the lowest position of the piston, and the third row shows the packings after the unloading.

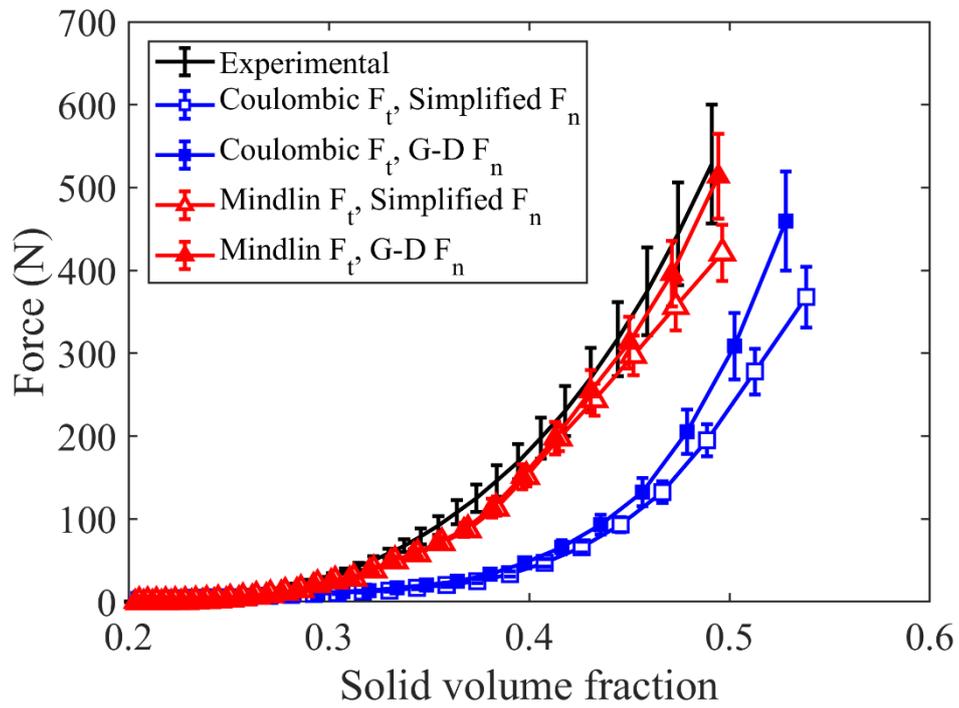

Figure 7. Force-solid volume fraction curves in the uniaxial compression of a packing of rubber cords: experimental results and DEM simulation results with different combinations of contact force models: Coulombic tangential force $F_t$ model, Mindlin tangential force $F_t$ models; simplified normal force $F_n$ model, and geometry-dependent (G-D) normal force $F_n$ model. The tests are replicated at least three times with different initial random configurations in both experiments and DEM simulations, with the scatter bars representing plus and minus a standard deviation.

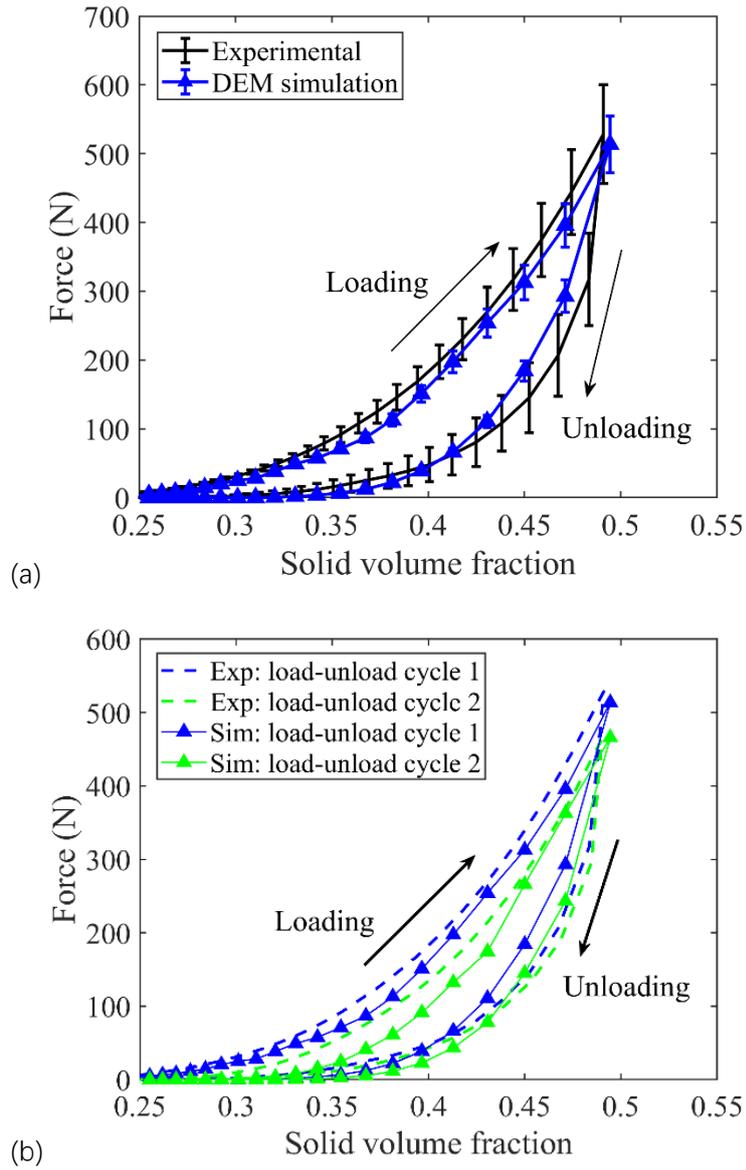

Figure 8. Comparison of experimental and DEM simulation results for the uniaxial compression tests of rubber cords: (a) load-unload cycle 1 with the scatter bars representing plus and minus a standard deviation; (b) load-unload cycles 1 and 2, and only the average values are shown. Each test is replicated at least three times with different initial random configurations in both experiments and DEM simulations.

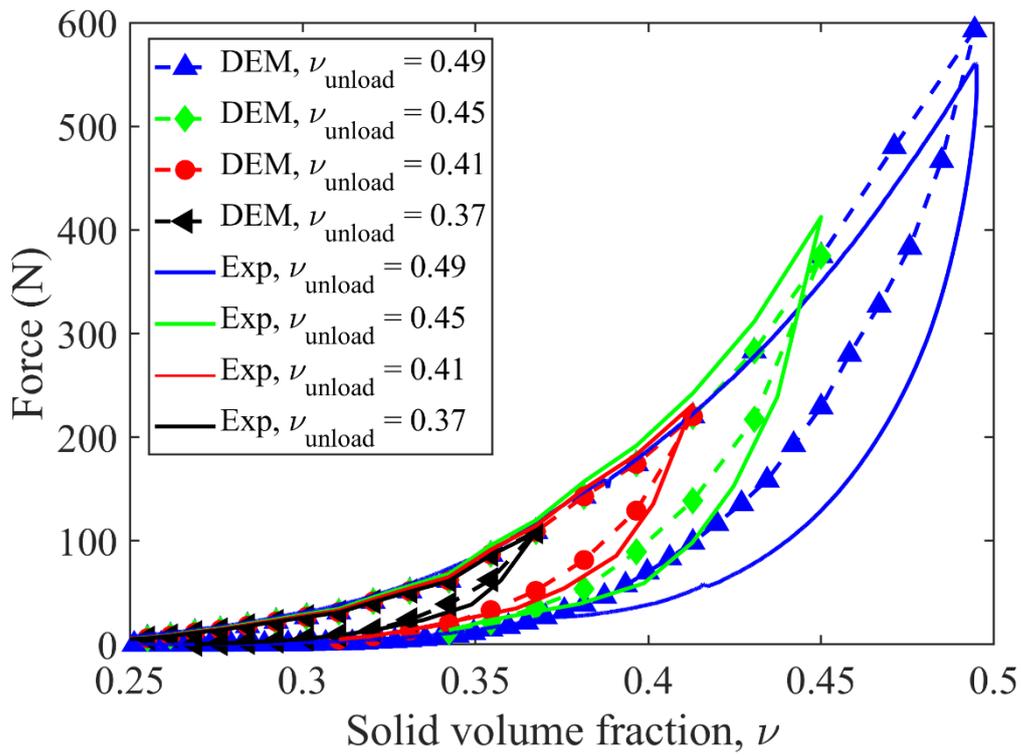

Figure 9. Comparison of experimental and DEM simulation results for the uniaxial compression tests of rubber cords: unloading at various solid volume fractions. Three tests are performed for each loading-unloading cycle, and the average results are presented in the plot. The scatter bars, which have similar sizes as those in Figure 8a, are deleted for visual clarity.

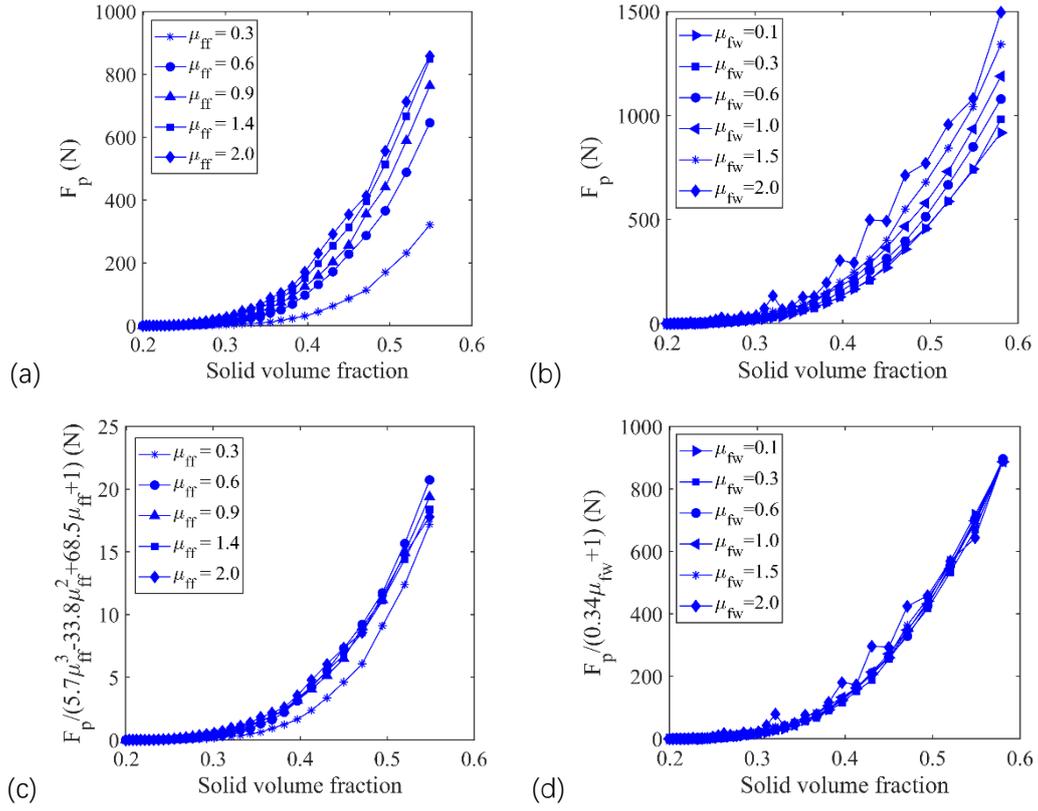

Figure 10. Effects of (a) fiber-fiber friction coefficient $\mu_{ff}$ and (b) fiber-wall friction coefficient $\mu_{fw}$ on the loading curves in the uniaxial compression tests. The $\mu_{ff}$ and $\mu_{fw}$ scaled load curves are plotted in (c) and (d), respectively. For the results in (a) and (c) $\mu_{fw}$ is equal to 0.6 and for those in (b) and (d) $\mu_{ff}$ is fixed at 1.4. Each test is replicated three times with different initial random configurations in the DEM simulations, and the average values are shown in these plots.

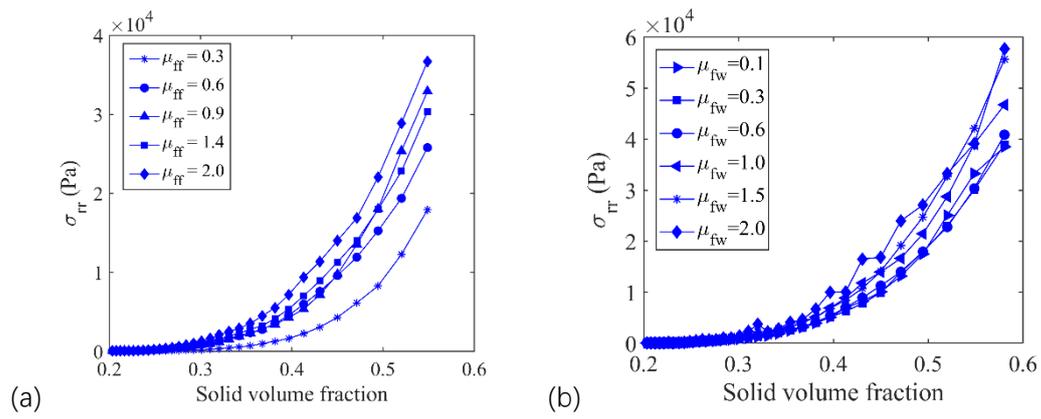

Figure 11. Effects of (a) fiber-fiber friction coefficient $\mu_{\text{ff}}$ and (b) fiber-wall friction coefficient $\mu_{\text{fw}}$ on the average normal stress $\sigma_{\text{rr}}$ exerted on the inner cylindrical wall of the container.

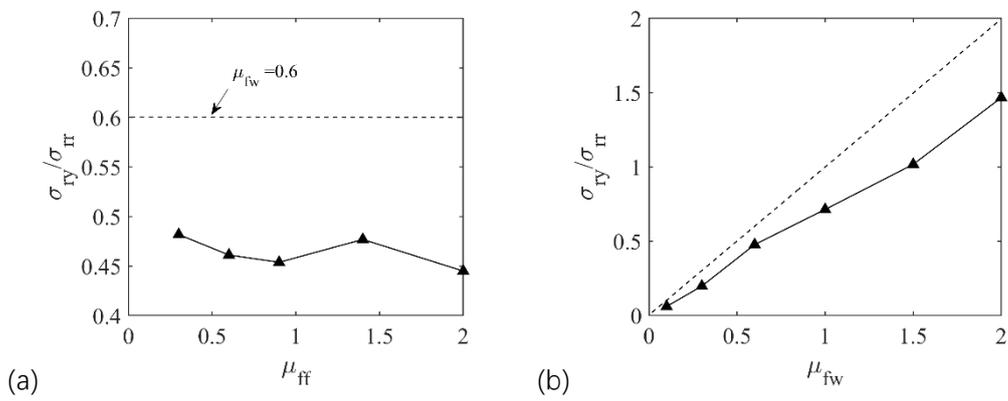

Figure 12. Effects of (a) fiber-fiber friction coefficient $\mu_{\text{ff}}$ and (b) fiber-wall friction coefficient $\mu_{\text{fw}}$ on the average shear-to-normal stress ratio exerted on the inner cylindrical wall of the container.

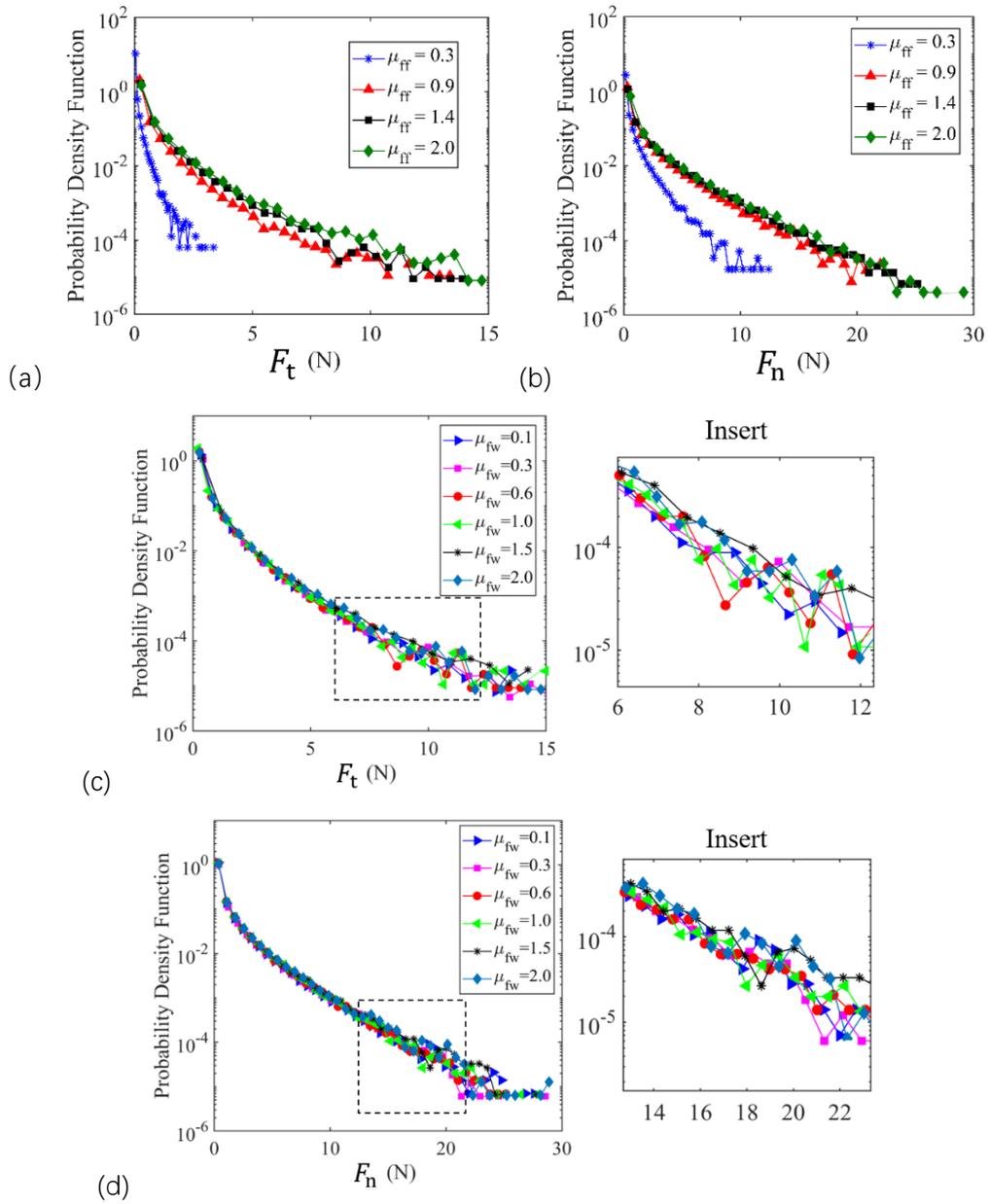

Figure 13. Effects of fiber-fiber friction coefficient $\mu_{ff}$ and fiber-wall friction coefficient $\mu_{fw}$ on the probability density functions of tangential and normal contact forces.

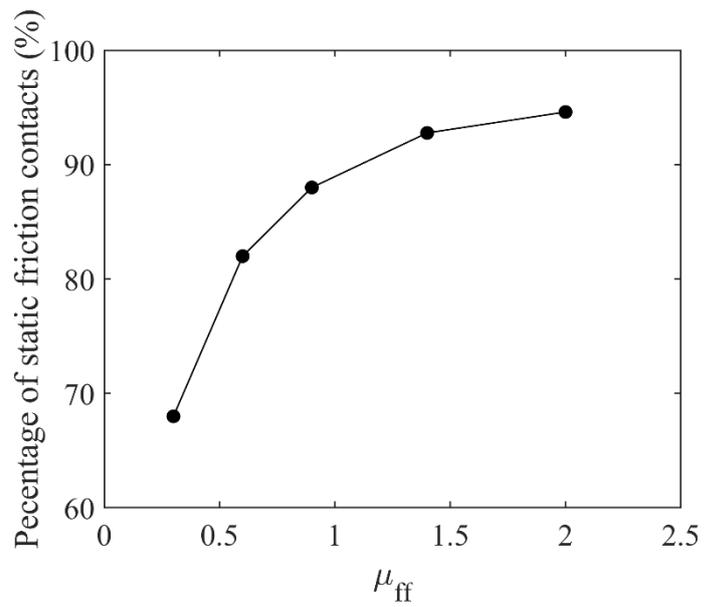

Figure 14. Percentages of static friction contacts with various fiber-fiber friction coefficients $\mu_{\text{ff}}$. The fiber-wall friction coefficient $\mu_{\text{fw}}$ is set to 0.6.